\documentclass[aps,prl,floatfix,twocolumn,superscriptaddress,showpacs,reprint]{revtex4-1}
\usepackage{graphicx}
\usepackage{mathrsfs}
\usepackage{amsmath}
\usepackage{subfigure}
\usepackage{bm}
\usepackage{verbatim}
\usepackage{color}
\usepackage{xcolor}
\usepackage{siunitx}
\usepackage{bbm}

\usepackage[colorlinks=true, pdfstartview=FitV,urlcolor=blue]{hyperref}

\begin{document}
\title{SU(3) Quantum Hall Ferromagnetism in SnTe}
\author{Xiao Li}\email{E-mail: lixiao@umd.edu}
\affiliation{Department of Physics, The University of Texas at Austin, Austin, Texas 78712, USA}
\affiliation{Condensed Matter Theory Center and Joint Quantum Institute, University of Maryland, College Park, Maryland 20742, USA}
\author{Fan Zhang}\email{E-mail: zhang@utdallas.edu}
\affiliation{Department of Physics, The University of Texas at Dallas, Richardson, Texas 75080, USA}
\author{A. H. MacDonald}
\affiliation{Department of Physics, The University of Texas at Austin, Austin, Texas 78712, USA}

\begin{abstract}
The $(111)$ surface of SnTe hosts one isotropic $\bar{\Gamma}$-centered and 
three degenerate anisotropic $\bar{M}$-centered Dirac surface states.  
We predict that a nematic phase with spontaneously broken $\mathcal{C}_3$ symmetry will occur in the presence of an external magnetic field when the $N=0 \; \bar{M}$ Landau levels are $1/3$ or $2/3$ filled.  
The nematic state phase boundary is controlled by a competition between intravalley Coulomb interactions that favor a valley-polarized state,
and weaker intervalley scattering processes that increase in relative strength with
magnetic field.  
An in-plane Zeeman field alters the phase diagram by lifting the three-fold $\bar{M}$ Landau level degeneracy, yielding a ground state energy with $2\pi/3$ periodicity as a function of Zeeman-field orientation angle.  
\end{abstract}
\maketitle

\begin{figure}[t]
\includegraphics[scale=0.77]{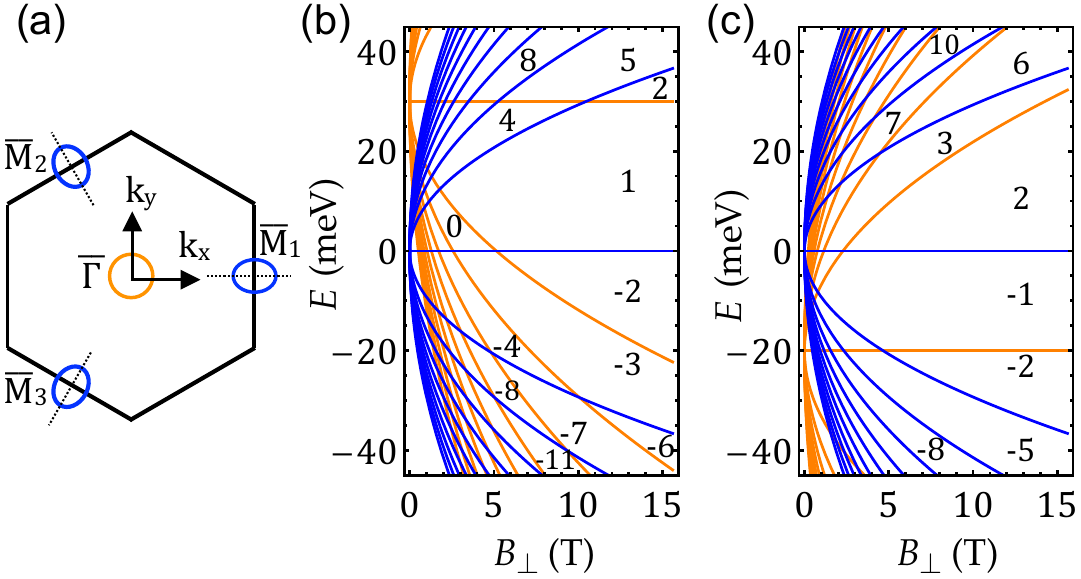}
\caption{(a) Typical equal-energy contours for gapless Dirac states on 
the $(111)$ surface of SnTe.
The three mirror invariance lines are indicated by dotted lines.
(b) and (c) Single-particle LL structures of (a).
The energy difference between $\bar{\Gamma}$ and $\bar{M}$ 
Dirac points is $30$~meV in (b) and $-20$~meV in (c).
Brown and blue lines represent the non-degenerate $\bar{\Gamma}$ LLs and the 
 three-fold degenerate $\bar{M}$ LLs, respectively. 
The integer labels in (b) and (c) give LL filling factors in spectral gaps.
\label{Fig:LL}}
\end{figure}

\indent{\color{cyan}{\em Introduction.}}---
Tin telluride (SnTe) is now attracting great attention
as the first topological insulator (TI) protected purely by crystalline symmetry.
Although its electronic band structure has been understood~\cite{SnTe} for decades,
the physical consequences of its band inversion have only recently been fully appreciated~\cite{TMI-Fu}.  
SnTe has a rocksalt crystal structure with two inter-penetrating face-centered cubic lattices,
and its bulk bands are inverted due to spin-orbit coupling at four $L$ points.
The mirror Chern number becomes nontrivial in mirror-invariant planes that each contains a pair of $L$ points.
Based on this property, Dirac surface states on selected surfaces respecting mirror symmetries
were first predicted~\cite{TMI-Fu} and later observed~\cite{TMI-exp1,TMI-exp2,TMI-exp3}.
The $(111)$ surface of SnTe~\cite{Buczko,Fu-111,FZ-SnTe,Ando-111,Duan,Feng,Safael,LiangFu-QSH} respects three 
mirror symmetries, and each protects an anisotropic gapless Dirac surface state at $\bar{M}$ and
a partner isotropic state at $\bar\Gamma$, as sketched in Fig.~\ref{Fig:LL}(a).


In crystalline topological insulators the top and bottom surfaces of 
thin films can be electrically isolated~\cite{FZ-SnTe} by 
breaking mirror symmetries on the side surfaces while leaving them time-reversal-invariant.
This behavior contrasts with the case of  strong TI thin films for which it is 
impossible to study single surface electrical properties because sidewalls can be gapped 
only by breaking time-reversal symmetry that generates Hall currents. 
The $(111)$ surface of SnTe therefore hosts a unique 
and relatively unexplored isolated two-dimensional electron gas (2DEG)
system in which the interplay between topological surface properties, valleytronics,
and many-body interactions is likely to yield unexpected phenomena. 

The integer quantum Hall (QH) effect is a hallmark of any 2DEG system.
When a 2DEG has Landau-level degeneracies due to spin, valley, 
and/or layer degrees of freedom~\cite{spin,valley,layer}, the interplay between Landau quantization and electron-electron interactions 
often leads to ground states in which symmetries associated with the aforementioned degrees of freedom are spontaneously broken.  Examples of broken symmetry states of this type, often referred to as 
QH ferromagnets, arise in GaAs and AlAs quantum wells~\cite{layer-exp,Shayegan}, 
single and multilayer graphene sheets~\cite{Yacoby,Kim,Lau}, 
and on the surfaces of silicon~\cite{Kane07,Kane14} and bismuth~\cite{Li,Zhu}.
In all instances of QH ferromagnetism studied to date, however, the noninteracting LL degeneracy $N$ has always been even. 
Thus one may wonder whether QH ferromagnetism with odd $N$ exists in some material, and how in this case the ground state breaks Hamiltonian symmetries.

In this Letter, we show that the $(111)$ surface of SnTe provides
a platform to explore $SU(3)$ QH ferromagnetism.
As illustrated in Fig.~\ref{Fig:LL}, the four Dirac cones on the $(111)$ surface give rise to four LL sequences.  The three $\bar{M}$ LL sequences are degenerate and related 
by a $\mathcal{C}_3$ rotational symmetry.  
The $\bar{\Gamma}$-$\bar{M}$ LL energy difference is not restricted
by any symmetry, and can~\cite{Feng} be tuned experimentally.
Under a uniform perpendicular magnetic field, these LL sequences 
are interwoven in a non-universal fashion.  We focus here on the case in 
which the $N=0$ $\mathcal{C}_3$ triplet is at the Fermi level,
well separated from all $\bar{\Gamma}$ LLs, 
and $1/3$ or $2/3$ filled to yield an integer total filling factor.  
We find that when only the stronger intravalley Coulomb interactions
are retained in the theory, valley symmetry is spontaneously broken to 
create a nematic state in which only one valley is occupied.
When the weaker intervalley scattering processes are retained, a 
broken translational symmetry state with coherence among three valleys appears 
beyond a critical magnetic field value.  
An in-plane Zeeman field couples to the 
nematic order parameter and influences the competition between broken rotational and 
translational symmetry states.

\indent{\color{cyan}{\em Surface state LLs.}}---
The SnTe $(111)$ Dirac surface states are described by the $\bm{k} \cdot \bm{p}$  
Hamiltonians~\cite{FZ-SnTe}:
\begin{eqnarray}
\!\!\!\!\!\!\!\!\!
H_{\bar{\Gamma}}= \hbar v(k_x s_y-k_y s_x), \, 
H_{\bar{M}_\lambda} = \hbar v_x k_x^{(\lambda)}s_y-v k_y^{(\lambda)}s_x,
\label{Eq:Hs}
\end{eqnarray}
where $v=4.40\times 10^{5}$\,{m/s} and $v_x=2.55\times 10^{5}$\,\text{m/s}~\cite{Feng}
are surface Fermi velocities, $\lambda=1,2,3$ labels the three inequivalent $\bar{M}$ valleys, 
and $\bm s$ is a surface Dirac pseudospin.
Microscopically the pseudospins~\cite{FZ-SnTe,FZ-TI} are valley-dependent 
linear combinations of spin and orbital operators that 
transform like spin under time reversal, spatial inversion, and mirror reflection.
$k_x^{(\lambda)}$ and $k_y^{(\lambda)}$ are explicitly defined in Fig.~\ref{Fig:LL}(a) for the $\bar{M}_1$ valley; 
local momentum-space coordinates in other valleys are obtained by 
appropriate $\mathcal{C}_3$ rotations.

In the presence of a uniform perpendicular magnetic field, the 2D kinetic momenta $\hbar\bm{k}$ in Eq.~\eqref{Eq:Hs}
are replaced by $\bm{\pi}=\hbar\bm{k}+e\bm{A}$, where $\bm{A}=(0,-B_{\perp} x)$.
The $\bar{\Gamma}$ LL energies are $E_{N,\pm}(\bar{\Gamma})$$=$$\pm\sqrt{2\hbar v^2 NeB_{\perp}}$,
reminiscent of the massless Dirac fermion LLs in graphene and on cleavage surfaces of $\rm Bi_2Se_3$.
Because the $\bar{M}$ surface states have anisotropic dispersions with valley-dependent 
orientations, we define valley-dependent raising operators
$a^{\dag}_{\lambda}$$=$$(\ell/\sqrt{2}\hbar)(\alpha_{\lambda} \pi_x-i\beta_{\lambda} \pi_y)$, where $\ell$$=$$\sqrt{\hbar/eB_\perp}$ is the magnetic length,  
$\alpha_{\lambda}$$=$$\eta\cos\theta_{\lambda}+i\eta^{-1}\sin\theta_{\lambda}$,
$\beta_{\lambda}$$=$$\eta^{-1}\cos\theta_{\lambda}+i\eta\sin\theta_{\lambda}$, $\eta$$=$$\sqrt{v_x/v}$,
and $\theta_{\lambda}=2(\lambda-1)\pi/3$.  With these definitions 
$H_{\bar{M}_\lambda}$ becomes
\begin{align}
H_{\bar{M}_\lambda} = \frac{\sqrt{2vv_x}\hbar}{\ell}
\begin{pmatrix}
0 & -ia^\dag_{\lambda} \\ ia_{\lambda} & 0
\end{pmatrix},
\end{align}
so that the  $\bar{M}$ LL energies and wavefunctions 
are $E_{N,\pm}(\bar{M}_\lambda)$$=$$\pm\sqrt{2\hbar v_x v N eB_\perp}$, 
and 
\begin{align}
\psi_{0\lambda} &=
\begin{pmatrix}
\phi_{0\lambda} \\ 0
\end{pmatrix},\;
\psi_{N>0,\lambda,\pm} = \frac{1}{\sqrt{2}}
\begin{pmatrix}
\phi_{N\lambda} \\ \pm\frac{\alpha^{\ast}_{\lambda}}{|\alpha_{\lambda}|}\phi_{N-1,\lambda}
\end{pmatrix},\label{wf}\\
\phi_{N\lambda} &= \mathcal{A}_{N\lambda}^{-1} e^{-\alpha_{\lambda}\beta_{\lambda}^{\ast}\xi_{\lambda}^2/2}H_N(\xi_{\lambda}).
\end{align}
Here $\mathcal{A}_{N\lambda}$$=$$\left({2^N N!\sqrt{\pi}|\alpha_{\lambda}|\ell}\right)^{\frac{1}{2}}$ is a normalization 
factor, $\xi_{\lambda}$$=$$(x-k_y\ell^2)/(|\alpha_{\lambda}|\ell)$, and $H_N(\xi)$ is the Hermite polynomial. 
Also note that $a\phi_N$$=$$-i\alpha^\ast/|\alpha|\sqrt{N}\phi_{N-1}$ and $a\phi_0$$=$$0$. In Fig.~\ref{Fig:LL} we plot LL spectra as a function of $B_\perp$ for cases with 
the $\bar{\Gamma}$ Dirac point energy above and below the
$\bar{M}$ Dirac point energy.
All the $\bar{\Gamma}$ LLs are non degenerate, whereas all the $\bar{M}$ LLs are threefold degenerate because of the $\mathcal{C}_3$ symmetry.

\indent{\color{cyan}{\em QH ferromagnetism in the $N$\,$=$\,$0$ triplet.}}---
We focus here on the case in which the $N\!=\!0$ LL triplet is $1/3$ or $2/3$ filled,
and ask whether Hamiltonian symmetries are spontaneously broken and whether broken symmetries give rise to charged excitation gaps which would yield an integer QH effect.
Because Coulomb interaction matrix elements are sensitive to the valley-dependent orientations of 
the anisotropic cyclotron orbits, the Hamiltonian is not invariant under rotations in the valley-space.
However, the small size of the momentum-space cyclotron orbits 
relative to their separation
implies that the number of electrons in each pocket is conserved; the only 
allowed large momentum transfer processes simply exchange electrons between valleys.  
Broken symmetry ground states are either Ising-like states in which the three symmetry equivalent 
valleys are occupied by different numbers of electrons, or XY-like states in which coherence 
is spontaneously established among three valleys
oriented along different directions.
The Ising-like state is a nematic one which 
lowers rotational symmetry, and the XY-like state is a commensurate charge-density-wave state which breaks the crystal translational symmetry.  
Interesting new physics is most likely to be experimentally accessible 
when the $N$\,$=$\,$0$ triplet is partially filled because of the  
large gap separating $N$\,$=$\,$0$ and $N \neq 0$ LLs in Dirac systems. After projecting to the $N$\,$=$\,$0$ triplet, states at 
$1/3$ and $2/3$ fillings are related by particle-hole symmetry within the triplet,
allowing us to focus on the $1/3$ case. In addition, we neglect the possibility of an accidental degeneracy between the $N\!=\!0$ $\bar{M}$
triplet and a $\bar{\Gamma}$ LL.

We employ the unrestricted Hartree-Fock (HF) approximation~\cite{Cote-AHM}
which is accurate~\cite{Schliemann-AHM} at the integer total filling factors of interest.
We therefore minimize the energy of Slater determinant trial wavefunctions
by solving self-consistent field equations with $3 \times 3$ mean-field Hamiltonians of the form:
\begin{align}
\mathcal{H}^{\rm HF}_{\lambda\sigma} &= E_0 \delta_{\lambda\sigma}
+Y_0^{\lambda\sigma}\Delta_{\sigma\lambda}(1-\delta_{\lambda\sigma}) \nonumber\\
&\qquad - X_0^{\lambda\sigma} \Delta_{\sigma\lambda}	
-\sum_{\tau\neq\lambda}Z_0^{\lambda\tau}\Delta_{\tau\tau}\delta_{\lambda\sigma},
\label{H-mf}
\end{align}
where $E_0$ is the single-particle LL energy which is an irrelevant constant that is subsequently set to $0$, and 
$\Delta_{\sigma\lambda}\!=\!\langle c^\dag_\sigma c_\lambda \rangle$ is the triplet density matrix.
In Eq.~\eqref{H-mf} $X_0^{\lambda\sigma}$, $Y_0^{\lambda\sigma}$, and $Z_0^{\lambda\tau}$
are respectively intravalley exchange, intervalley Hartree, and intervalley exchange integrals.  
We use an envelope function approximation for intravalley processes,
which are enhanced by the long-range tail of the Coulomb interactions and therefore dominant, and approximate 
intervalley processes using a phenomenological interaction constant  
$U \sim 2\pi e^2/\epsilon K$ where ${\bm K}$ is a primitive reciprocal lattice vector.
It follows that the Hartree integral is $Y_0^{\lambda\sigma}\!=\! (2\pi\ell^2)^{-1} UF_{00}^{\lambda\sigma}(0) F_{00}^{\sigma\lambda}(0)$, 
and that the exchange integrals are 
\allowdisplaybreaks[4]
\begin{eqnarray}
	X_0^{\lambda\sigma}&=&\int  \frac{d^2 {\bm k}}{(2\pi)^2}\frac{2\pi e^2}{\epsilon k}
		F_{00}^{\lambda\lambda}(\bm{k}) F_{00}^{\sigma\sigma}(-\bm{k})
		e^{i k_x k_y \ell^2 \mathcal{W}_{X}^{\lambda\sigma}}, \\
	Z_0^{\lambda\sigma}&=& U \int \frac{d^2 {\bm k}}{(2\pi)^2} 
		F_{00}^{\lambda\sigma}(\bm{k}) F_{00}^{\sigma\lambda}(-\bm{k})
		e^{i k_x k_y \ell^2 \mathcal{W}_{Z}^{\lambda\sigma}}, \label{Eq:Zterm}
\end{eqnarray}
where $\epsilon= (\epsilon_\text{SnTe}+1)/2\sim 20$~\cite{SnTe-Dielectric}  
is the effective dielectric constant,
and $F_{00}^{\lambda\sigma}(\bm{k})$ is a form factor that accounts for the system's valley-dependent 
cyclotron-orbit shape:
\begin{eqnarray}
\!\!\!\!\!
	F_{00}^{\lambda\sigma}(\bm{k}) = \dfrac{\sqrt{2}}{\sqrt{|\alpha_{\lambda}||\alpha_{\sigma}|(\gamma_{\lambda}+\gamma_{\sigma}^{\ast})}}
	\exp\left[ \frac{(k_x^2+\gamma_{\lambda}\gamma_{\sigma}^{\ast}k_y^2)\ell^2}{-2(\gamma_{\lambda}+\gamma_{\sigma}^{\ast})} \right].  \label{Eq:FF}
\end{eqnarray}

In the above integrals $\mathcal{W}_{X}^{\lambda\sigma}\!=\!1-w_{\lambda\lambda}-w_{\sigma\sigma}$ and
$\mathcal{W}_{Z}^{\lambda\sigma}\!=\!1-w_{\lambda\sigma}-w_{\sigma\lambda}$,
with $w_{\lambda\sigma}\!=\!\gamma_\sigma^\ast/(\gamma_\lambda+\gamma_\sigma^\ast)$ and
$\gamma_{\lambda}\!=\!\beta_{\lambda}/\alpha_{\lambda}$.
If the surface states were isotropic (i.e., $v\!=\!v_x$ and $\gamma_\lambda\!=\!1$),
$F_{00}^{\lambda\sigma}(\bm{k})$ would reduce to the circular cyclotron orbit 
form-factor $\exp(-k^2\ell^2/4)$~\cite{Cote-AHM} that accounts for the quantum-smeared $N=0$ orbit size and shape.  The corrections in Eq.~\eqref{Eq:FF} account for the anisotropy of the triplet 
cyclotron-orbits, and for the $2\pi/3$ differences in anisotropy orientation 
illustrated in Fig.~\ref{Fig:LL}, which play an essential role in the interaction physics. 
Because of the $\mathcal{C}_3$ symmetry, the intravalley 
exchange integral matrix $X_0^{\lambda\sigma}$ 
only has two inequivalent matrix elements, larger exchange integrals for electrons in the same valley ($X_0^S$) 
on its diagonal and smaller exchange integrals for electrons in different valleys ($X_0^D$) for its off-diagonal elements.
Because we take the interaction responsible for valley-exchange electron-electron scattering to 
be short-range, the intervalley integrals have only off-diagonal matrix elements all of which 
have the same value ($Y_0$ and $Z_0$).  

The broken symmetry ground state minimizes the total energy with respect 
to the five parameters that characterize the valley spinor,
 $(r_1e^{i\varphi_1}, r_2e^{i\varphi_2}, r_3 )^{\rm T}$,
shared by all LL orbitals. 
Up to a spinor-independent constant, the energy per electron is
\begin{align}
\mathcal{E}=2[(X_0^S\!-\!X_0^D)\!-\!(Z_0\!-\!Y_0) ] \; (r_1^2 r_2^2+r_2^2 r_3^2+r_3^2 r_1^2).
\label{EF}
\end{align}
The $1/3$ filling ground state therefore depends on the sign of
a linear combination of the interaction parameters calculated above. 
The energy is independent of $\varphi_1$ and $\varphi_2$ because   
of separate particle number conservation in each valley.  
The spinor-dependent factor on the right hand side of Eq.~\eqref{EF}
reaches its minimum value $0$ when the spinor is a single-valley state
$(r_1,r_2,r_3)=(1,0,0)$, $(0,1,0)$, or $(0,0,1)$ and its maximum value 
$1/3$ when the ground state is an equal-weight three-valley state, $(r_1,r_2,r_3)=(1,1,1)/\sqrt{3}$.

Exchange energies are always stronger between orbitals that are more similar.
Accordingly, the exchange integrals between electrons in the same valley are 
stronger than those between electrons in different valleys ($X_0^S\!>\!X_0^D\!>\!0$) 
because of the difference in cyclotron orbit orientations.
It follows that the ground state is completely valley polarized 
unless intervalley scattering plays a role. 
LL interaction physics in SnTe surface 2DEGs therefore contrasts strongly 
with the case of graphene 2DEGs which has identical isotropic Dirac cones in
two different valleys, implying that $X_0^S\!=\!X_0^D$.  Broken valley
symmetry states at $\nu\!=\!\pm1$ in graphene~\cite{DasSarma,Goerbig,Alicea,Nomura} therefore have Heisenberg character when valley-exchange processes are neglected.

For the relatively modest anisotropy parameter $\eta\sim0.75$ of SnTe we find that
the difference between the same-valley and different-valley exchange energies is 
small, $X_0^S\!-\!X_0^D$\,$=$\,$0.0541\,e^2/(\epsilon\ell)$. The short-range
valley-exchange scattering processes are approximated by a momentum-independent  interaction with strength $U$ we find that $Z_0 - Y_0$ is positive and scales as $U/\ell^2$. This allows the weak valley-exchange scattering to play a role at stronger fields, 
favoring a ground state which has coherence between all three valleys, and is therefore a charge-density-wave state with broken translational symmetry.  
For $U=2\pi e^2/\epsilon K$ = 0.85 eV$\cdot$nm$^2$, a first-order quantum phase transition between nematic valley-polarized and 
valley-coherent charge-density-wave states occurs at $B_{\perp}^c\simeq \SI{11}{T}$.  
In graphene~\cite{DasSarma,Goerbig,Alicea,Nomura} and 
monolayer MoS$_{2}$~\cite{Xiao-MoS2}, however, there is no competition with valley-dependent exchange, and the same mechanism induces a charge-density-wave ground state
at all field strengths at filling factors $\nu\!=\!\pm1$.  

\begin{figure}[t]
\includegraphics[scale=0.63]{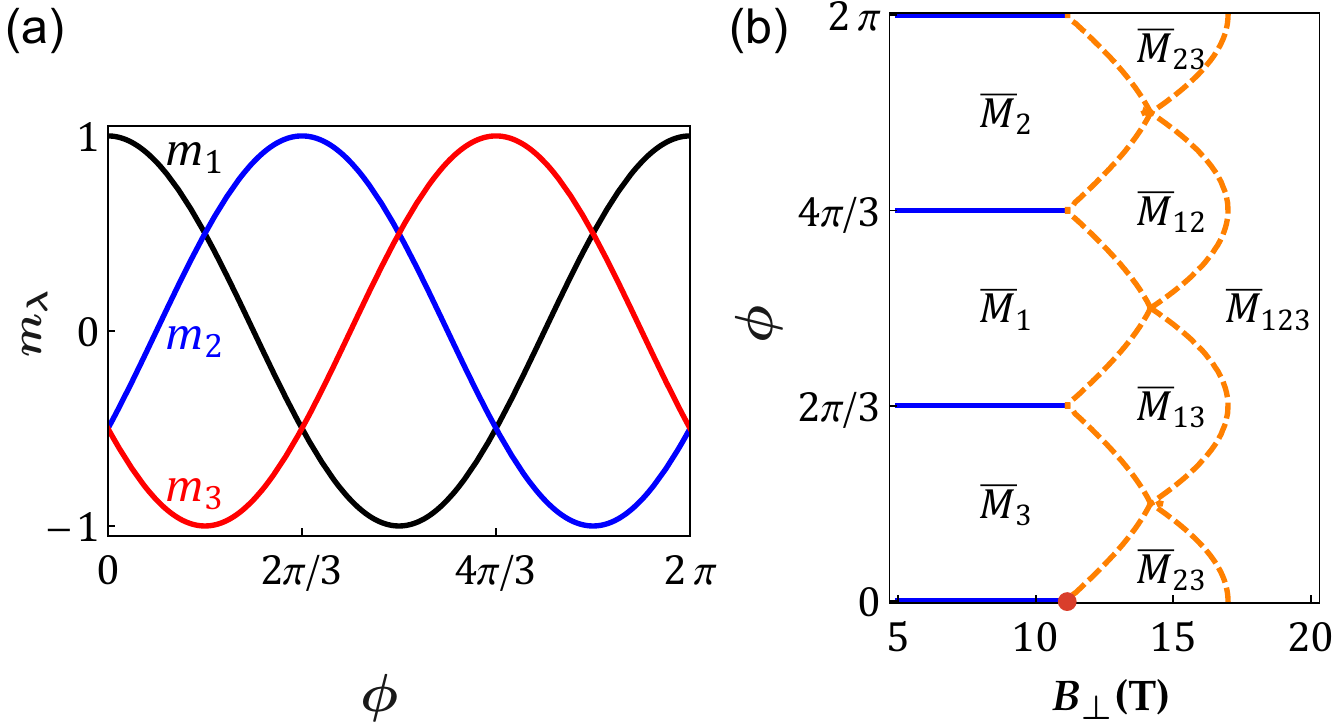}
\caption{(a) In-plane Zeeman energies $m_\lambda$ 
in units of $\sqrt{2}\eta^2 g\mu_{B}B_{\parallel}/9$ as a function of $B_{\parallel}$ orientation.
(b) Phase diagram for the state at $1/3$ filling of the $N=0$ LL triplet for $\epsilon g B_{\parallel}\!=\!{300}$~T
and $U = \SI{0.85}{eV$\cdot$nm$^2$} $. 
The red dot denotes the critical field $B_{\perp}^c$
at which a first-order transition occurs between a valley polarized and a three valley symmetric
state occurs at $B_{\parallel}\!=\!0$.  The solid blue lines are first-order phase boundaries between complete valley polarized 
states, and the dashed red lines are continuous transition boundaries between 
states with coherence between different numbers of valleys.  The phases
labeled by $\bar{M}_i$, $\bar{M}_{ij}$, and $\bar{M}_{123}$ have a full LL spinor that is a coherent superposition of components involving one, two, and three valleys, respectively.  
\label{Fig:PD}}
\end{figure}

\indent{\color{cyan}{\em Zeeman field effects.}}---
We have so far neglected Zeeman coupling, 
which greatly enriches the interaction induced integer QH effect of 
SnTe. We write the total magnetic field as 
\begin{align}
	\bm{B} = B_{\parallel}(\cos\phi\,\hat{x}+\sin\phi\,\hat{y})+ B_{\perp}\,\hat{z},
\label{Eq:Bfield}
\end{align}
using the coordinate frame defined in Fig.~\ref{Fig:LL}(a) for the crucial
in-plane-field orientation $\phi$.  For general $\phi$ the Zeeman field breaks mirror symmetries and
couples to the order parameter by producing valley-dependent single-particle energies~\cite{FZ-SnTe,FZ-Zeeman}, i.e., $E_0 \to E_0 + m_\lambda$ in the HF Hamiltonian~\eqref{H-mf} with 
\begin{align}
	m_{\lambda}=\frac{1}{2}\alpha g\mu_{B}\left[\frac{2\sqrt{2}}{3}B_{\parallel}\cos(\phi-\theta_{\lambda})+\frac{1}{3}B_{\perp}\right], 
\end{align}
where $g$ is an electron $g$-factor, 
$\mu_{B}$ is the Bohr magneton, and $\alpha\!=\!\eta^2/3$ is the 
real spin weight of the surface pseudospin~\cite{FZ-SnTe,FZ-Zeeman}.
The perpendicular field $B_\perp$ does not break the $\mathcal{C}_3$ symmetry and
contributes only an irrelevant~\cite{GM-Zeeman} valley-independent energy shift of the $N\!=\!0$ triplet.
In contrast, the in-plane field $B_\parallel$ breaks 
$\mathcal{C}_3$ symmetry and lifts the $N\!=\!0$ triplet degeneracy.
It follows that in-plane fields can yield an integer QH effect at $1/3$ and $2/3$ filling of the $N\!=\!0$ triplet 
even in the absence of interactions, as illustrated in Fig.~\ref{Fig:PD}(a).
When $\hat{B}_\parallel$ is parallel (anti-parallel) to $\bar{\Gamma}$-$\bar{M}$, the
triplet level degeneracy is reduced to a two-fold degeneracy
at $1/3$ ($2/3$) filling.  
When $\hat{B}_\parallel$ is (anti)parallel to $\bar{\Gamma}$-$\bar{K}$, the single-particle
gaps at $1/3$ and $2/3$ filling are non-zero and identical.

Valley dependent Zeeman coupling competes with electron-electron interactions, 
and greatly enriches the phase diagram by adding 
$\delta\mathcal{E}\! \to \!\sum_{\lambda}m_\lambda r_\lambda^2$ to Eq.~\eqref{EF},
favoring complete valley polarized states.  
The phase diagram at $\epsilon g B_\parallel\!=\!\SI{300}{T}$ is illustrated in Fig.~\ref{Fig:PD}(b).
For $B_\perp \!<\! B_\perp^c \!=\! (\SI{2.81}{eV$\cdot$nm$^2$}/{U})^2$,
interactions prefer a valley polarized state and $\phi$ simply selects which valley is occupied.
First-order phase transitions occur at $\phi\!=\!\theta_\lambda$.
When Zeeman coupling to a parallel field is included, the abrupt transition from valley-polarized to three-valley coherent states 
is interrupted by a region in which $\phi$-dependent two-valley coherent states are stable. 
The stability range of the two-valley coherent state is widest when 
$\phi = \theta_\lambda$.  Finally,
when $B_\perp$ is further increased, 
three-valley coherent states finally 
emerge, but with $\phi$-dependent and unequal valley populations.   
Valley coherence can therefore be modified and continuously tuned by the in-plane Zeeman field.

The shape of the phase diagram in Fig.~\ref{Fig:PD}(b) is only weakly dependent on the 
numerical value of $\epsilon g B_\parallel$.  
The three first-order transition lines and $B_\perp^c$ are independent of $\epsilon g B_\parallel$ changes.
A larger value of $\epsilon g B_\parallel$ expands the areas with two-valley coherent states 
to larger $B_\perp$.
Stronger short range interactions shift $B_\perp^c$ to smaller values because 
intervalley interactions increase in importance.  
On the other hand larger surface state anisotropy would 
increase the critical perpendicular field $B_\perp^c$.

\indent{\color{cyan}\em Discussion.}---
We have shown that because of valley-dependent anisotropic cyclotron orbits, 
intravalley electron-electron interactions in SnTe can reduce rotational symmetries 
and lift the three-fold degeneracy of the $\bar{M}$ valley $N=0$ LLs.  
The physics which drives this broken symmetry 
is similar to that~\cite{Sondhi10,Sondhi13} responsible for valley 
polarized nematic states in parabolic spinful band systems
with an even number of valleys.  The triplet case discussed here 
is distinguished by its $SU(3)$ order parameter space, and by the way Zeeman interactions
with parallel fields couple to the order parameter.  Zeeman interactions play a key role because 
parallel fields break the mirror and $\mathcal{C}_3$ symmetries 
that protect the three $\bar{M}$ valley surface Dirac states.  
Our theory can be extended to study QH effects on the surfaces of silicon~\cite{Kane07,Kane14} and bismuth~\cite{Li,Zhu}.
We predict that, intervalley interactions will become important at sufficiently strong fields and drive a transition from a valley-polarized nematic state to a commensurate 
charge-density-wave state with intervalley coherence, and that the phase boundary between
them is enriched by in-plane field Zeeman coupling.    
%
%
%
%
%

In the thin-film case~\cite{STI}, Coulomb interactions between 2DEGs
localized on top and bottom surfaces are important.  
We expect that in this case integer QH effects at total filling factor $\nu_{T}=0$ will be spatially indirect exciton condensates. 
Our theory can also be extended to study $SU(3)$ symmetry breaking in $N\!\neq\!0$ LL triplets.
In this case, we expect similar competing ordered states but smaller exchange interactions, and the Zeeman-coupling effect is enriched but will still retain the 
$2\pi/3$ periodicity in $B_\parallel$ orientations.

The first step experimentally in studying the interaction physics we address would be to verify the LL
structure we predict using field-angle dependent magnetoresistance or magnetic torque magnetometry~\cite{Li,Zhu}. 
The energies of $\bar{\Gamma}$ and $\bar{M}$ valley LLs are $\sqrt{Nv^2B_\perp}$ and $\sqrt{Nvv_xB_\perp}$, respectively.
Their crossings, illustrated in Fig.~\ref{Fig:LL}, should lead to pronounced peaks in longitudinal magnetoresistance.  
We note the LL crossing fields may be controllable by varying the surface potential~\cite{Feng}, which tunes the energy difference between the $\bar{\Gamma}$ and $\bar{M}$ Dirac points.
An in-plane Zeeman field splits the $SU(3)$-invariant triplets, with $2\pi/3$ periodicity as a function of 
Zeeman-field orientation.   
Since Shubnikov-de Haas oscillations have recently been observed on the $(001)$ surface of 
SnTe~\cite{Madhavan}, we expect future progress to be rapid.  
The  phase diagram Fig.~\ref{Fig:PD}(b) is expected to be observable only in low-disorder samples,
since the transport activation gaps associated with broken symmetry states are of the order of $e^2/(\epsilon\ell)\sim 56\sqrt{B_{\perp}~[\mbox{T}]}/\epsilon$~meV.  The collective modes~\cite{G-mode} of valley coherent states are expected to be gapless,
while those of valley polarized states are expected to be gapped. 

\indent{\color{cyan}\em Acknowledgments.}---
We thank Qian Niu for helpful discussions. 
X.L. was supported by DOE (DE-FG03-02ER45958, Division of Materials Science and Engineering) and Welch Foundation (F-1255) in Austin, and is currently supported by NSF-JQI-PFC and LSP-CMTC in Maryland.
X.L. also acknowledges discussions with Fengcheng~Wu, Rohit~Hegde and Ming~Xie in Austin.  
F.Z. is supported by UT Dallas research enhancement funds. F.Z. also thanks the Aspen Center for Physics (the NSF Grant \#1066293 and the Trustee’s Fund) for hospitality during the finalization of this work.
A.H.M. is supported by DOE Division of Materials Sciences and Engineering grant DE-FG03-02ER45958, and by Welch foundation grant TBF1473. 
 
\bibliographystyle{apsrev4-1}


\begin{thebibliography}{100}
\bibitem{GM-Zeeman}
Note that the Zeeman effects of $B_\perp$ are distinct~\cite{FZ-SnTe} for $\bar{\Gamma}$ and $\bar{M}$ states.
\end{thebibliography}

\end{document}